\documentclass[prl,twocolumn,showpacs,preprintnumbers,amsmath,amssymb]{revtex4}
\usepackage{amsmath,amssymb}
\usepackage{graphicx}% Include figure files
\usepackage{dcolumn}% Align table columns on decimal point
\usepackage{bm}% bold math
\usepackage{color}
\usepackage{epstopdf}
\newcommand{\mr}{\mathrm}

\begin{document}
\title{Hyper-Ramsey Spectroscopy of Optical Clock Transitions}

\author{V. I. Yudin and A. V. Taichenachev}
\affiliation{Institute of Laser Physics SB RAS, Novosibirsk 630090, Russia}
\affiliation{Novosibirsk State University, Novosibirsk 630090, Russia}
\affiliation{Novosibirsk State Technical University, Novosibirsk 630092, Russia}
\author{C. W. Oates, Z. W. Barber, N. D. Lemke, and A. D. Ludlow}
\affiliation{National Institute of Standards and
Technology, 325 Broadway, Boulder, CO 80305, USA}
\author{U. Sterr, Ch. Lisdat, and F. Riehle}
\affiliation{Physikalisch-Technische Bundesanstalt (PTB), Bundesallee 100, 38116 Braunschweig, Germany}
\date{\today}

\begin{abstract}
We present non-standard optical Ramsey schemes that use pulses
individually tailored in duration, phase, and frequency to cancel
spurious frequency shifts related to the excitation itself. In
particular, the field shifts and their uncertainties of Ramsey fringes can be radically
suppressed (by 2-4 orders of magnitude) in comparison with the usual Ramsey
method (using two equal pulses) as well as with single-pulse Rabi
spectroscopy. Atom interferometers and optical clocks based on
two-photon transitions, heavily forbidden transitions or magnetically induced spectroscopy could significantly benefit from
this method. In the latter case these frequency shifts can be
suppressed considerably below a fractional level of 10$^{-17}$.
Moreover, our approach opens the door for the high-precision optical
clocks based on direct frequency comb spectroscopy.
\end{abstract}


\maketitle

Presently laser spectroscopy and fundamental metrology are among
of the most important and actively developed directions in modern
physics. Frequency and time are the most precisely measured
physical quantities, which, apart from practical applications
(in navigation and information systems), play critical roles in tests of fundamental physical theories (such as QED, QCD,
unification theories, cosmology etc.) \cite{did04}. Now laser
metrology is confronting the challenging task of creating an optical clock with
fractional inaccuracy and instability at the level of
$10^{-17}$-$10^{-18}$. Indeed, considerable progress has
already been achieved along this path for both ion-trap \cite{rozenblad08} and
atomic lattice-based \cite{aka08,ludlow08} clocks.

Work in this direction has stimulated the development of novel
spectroscopic methods (e.g., spectroscopy using quantum logic
\cite{schmidt05} and magnetically induced spectroscopy
\cite{bar06}).  For some of the promising clock systems, one of the key
unsolved problems is the frequency
shift of the clock transition due to the excitation pulses
themselves. For the case of magnetically induced spectroscopy these
shifts (quadratic Zeeman and ac-Stark shifts) could ultimately limit the
achievable performance. Moreover, for ultra-narrow transitions (e.g., electric octupole \cite{hos09}
and two-photon transitions \cite{fis04,badr06}) the ac-Stark shift can
be so large in some cases to rule out high accuracy clock performance at all.   A similar limitation
exists for clocks based on direct frequency comb spectroscopy \cite{fortier06,stowe08} due to ac-Stark shifts induced by large numbers of off-resonant laser modes.

In this paper we propose a general solution to this important
problem, which is based on the development and generalization of
the Ramsey method \cite{ram50}. We have found that in contrast
with the single-pulse (Rabi) technique, multi-pulse Ramsey
spectroscopy offers several ways (e.g., pulse durations,
frequencies, and phases) to significantly manipulate (due to
interference effects) induced frequency shifts of the spectroscopy
signals. In particular, for special excitation schemes (which we
refer to as ``hyper-Ramsey'') the resulting ac-Stark shift depends
on the laser intensity in an essentially non-linear way. Such
unusual and unexpected behavior allows us to dramatically suppress
these shifts and their uncertainties (most critical
for clocks) by two to four orders of magnitude with strongly
relaxed control requirements for the experimental parameters.
Additionally, we have found that these schemes can have a greatly
reduced sensitivity to the pulse areas, which makes the procedure
robust and accessible experimentally. Thus, this method can be
readily implemented (as needed) in a variety of existing and
proposed clock systems
\cite{rozenblad08,aka08,ludlow08,schmidt05,bar06,hos09,fis04,badr06,fortier06,stowe08,zan06a}.
Our approach could lead to significant progress for atomic clocks:
it will improve several key existing optical clock systems, and
could enable new systems that were not previously thought to be
competitive.

Note, some variants of the Ramsey technique were proposed
in \cite{zan06a,tai09} to cancel the overall field shift. However, the
shift uncertainties (caused by the fluctuations of field
parameters) were not significantly reduced in \cite{zan06a,tai09}, as it is possible with the
hyper-Ramsey method.

\begin{figure}[t]
\footnotesize
\includegraphics[width=8cm]{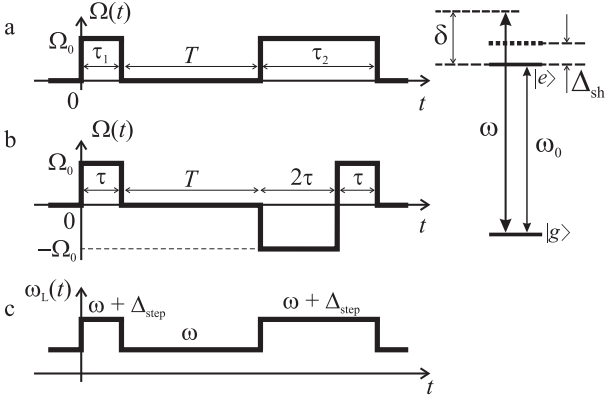}
\caption{Ramsey pulses with Rabi frequency $\Omega_0$ of different duration ($\tau_1$ and $\tau_2$; panel (a)) and with a phase step in the second pulse ($\tau_2 = 3\tau_1$; panel (b)). During the pulses, we step the laser frequency $\omega$ by $\Delta_{\rm step}$ (panel (c) and text). Also shown is a two-level atom with splitting $\omega_0$, detuning $\delta$ of the laser with frequency $\omega$ during dark time $T$, and excitation related shift $\Delta_\mr{sh}$ during pulses. }
\label{fig:scheme}
\end{figure}

The hyper-Ramsey spectroscopy schemes (Fig.~\ref{fig:scheme}) are
based on time-separated pulses that can have different durations,
frequencies, and phases. The action of a single light pulse (with
frequency $\omega_\mr{p}$,  duration $\tau$, and Rabi frequency
$\Omega_0$) on two-level atoms with ground and excited states,
$|g\rangle$$=$${\small \left(\begin{array}{c} 0 \\ 1
\end{array}\right)}$ and $|e\rangle$$=$${\small \left(\begin{array}{c} 1 \\
0 \end{array}\right)}$ (separated by the unperturbed energy $\hbar \omega_0$), is described by the matrix:
\begin{eqnarray}\label{W}
&&\widehat{W}(\tau,\Omega_0,\delta_\mr{p})=\\
&&\begin{pmatrix}
    \cos\left(\frac{\Omega \tau}{2}\right)
                +\frac{i\delta_\mr{p}}{\Omega}\sin\left(\frac{\Omega \tau}{2}\right)&

    \frac{i\Omega_0}{\Omega}\sin\left(\frac{\Omega \tau}{2}\right) \\
    \frac{i\Omega_0}{\Omega}\sin\left(\frac{\Omega \tau}{2}\right) &
    \cos\left(\frac{\Omega \tau}{2}\right)
                -\frac{i\delta_\mr{p}}{\Omega}\sin\left(\frac{\Omega \tau}{2}\right)
\end{pmatrix},\nonumber
\end{eqnarray}
where $\Omega=\sqrt{\Omega_0^2+\delta_\mr{p}^2}$ is the generalized
Rabi frequency. The detuning during pulse
$\delta_\mr{p}=\omega_\mr{p}-\omega_0-\Delta_\mr{sh}$ contains the
excitation related shift $\Delta_\mr{sh}$ (see
Fig.~\ref{fig:scheme}, level scheme) due to the influence of other
(far-off-resonant) transitions. Within the frequency interval corresponding to the narrow clock resonance the variation of $\Delta_\mr{sh}$ on $\omega_\mr{p}$ is negligible, i.e., $\Delta_\mr{sh}$ is a constant (for fixed $\Omega_0$).

During the dark period
between the pulses, excitation related shifts (which produce
the total actual shift $\Delta_\mr{sh}$) are absent (e.g., the ac-Stark shift
from the laser) or can be turned off (like the Zeeman shift). If
during the dark period $T$ the laser frequency is $\omega$, then
the free evolution is described by the matrix with detuning $\delta=\omega-\omega_0$:
\begin{equation}\label{V}
\widehat{V}(T\delta)=
\left(\begin{array}{cc}
e^{iT\delta /2}&0\\
0&e^{-iT\delta /2}
\end{array}\right).
\end{equation}
In the general case, the laser frequency during the
pulse does not have to be the same as the frequency during the
dark time, i.e., $\omega_\mr{p} \neq\omega$ \cite{tai09}.  As we will see,
at times it can be useful to approximately offset the induced
shift, $\Delta_\mr{sh}$, by stepping the laser frequency only
during the pulses by a fixed $\Delta_\mr{step}$, i.e.,
$\omega_\mr{p} = \omega + \Delta_\mr{step}$ (see
Fig.~\ref{fig:scheme}c).  Thus, in the general case the detuning
during the pulses can be written as $\delta_\mr{p} = \delta -
\Delta$, where $\Delta = \Delta_\mr{sh} - \Delta_\mr{step}$ is the effective frequency shift (during the pulse).

If at $t=0$ atoms are in the lower level $|g\rangle$, then after
the action of two pulses of duration $\tau_1$ and $\tau_2$
separated by dark period $T$ (see Fig.~1a) the population
$n_\mr{e}$ of atoms in the excited state is determined by
\begin{equation}\label{eq:n_e}
n_\mr{e}=\left|\langle e|
\widehat{W}(\tau_2,\Omega_0,\delta -\Delta)
\widehat{V}(T\delta)
\widehat{W}(\tau_1,\Omega_0,\delta-\Delta)
|g\rangle \right|^2.
\end{equation}
Equation~(\ref{eq:n_e}) describes Ramsey fringes (as a function of
variable detuning $\delta$, but with fixed $\Delta$). The presence
of the additional shift $\Delta$ in the course of the pulse action
leads to a shift $\overline{\delta\omega}_0$ of the position (top or
bottom) of the central Ramsey fringe with respect to the unperturbed
frequency $\omega_0$. To investigate the dependence of
$\overline{\delta\omega}_0$ on $\Delta$ we present the signal
$n_\mr{e}$ as a Taylor expansion in terms of the dimensionless
parameter ($T\delta$):
\begin{equation}
n_\mr{e}=a^{(0)}+a^{(1)}(T\delta)+a^{(2)}(T\delta)^2+... .\,.
\label{eq:n_e_d}
\end{equation}
The coefficients $a^{(j)}$ are expanded in the powers of $\Delta/\Omega_0$:
\begin{eqnarray}
a^{(0)}&=&{\cal A}_{0}^{(0)}+{\cal A}_{2}^{(0)}(\Delta/\Omega_0)^2+{\cal A}_{4}^{(0)}(\Delta/\Omega_0)^4+...\nonumber\\
a^{(1)}&=&{\cal A}_{1}^{(1)}(\Delta/\Omega_0)+{\cal A}_{3}^{(1)}(\Delta/\Omega_0)^3+...\nonumber\\
a^{(2)}&=&{\cal A}_{0}^{(2)}+{\cal A}_{2}^{(2)}(\Delta/\Omega_0)^2+{\cal A}_{4}^{(2)}(\Delta/\Omega_0)^4+...
\label{eq:Aj}
\end{eqnarray}
The occurrence of terms with only odd or even powers is the direct
consequence of the symmetry of Eq.~(\ref{eq:n_e}) that does not
change under the simultaneous substitutions $\delta \to -\delta$ and
$\Delta \to -\Delta$, which holds for any sequence of pulses
described by the matrices $\widehat{W}$ and $\widehat{V}$.

Even if the actual level shift $\Delta_\mr{sh}$ is comparable to or
larger than $\Omega_0$, we can always apply a frequency step
$\Delta_{\rm step}$ (e.g., with an acousto-optic modulator) during
excitation to achieve the condition $|\Delta/\Omega_0| \ll 1$ for an
effective shift $\Delta$. $\Delta_{\rm step}$ can be evaluated
experimentally by variation of the dark period $T$. If $\Delta_{\rm
step} \neq \Delta_\mr{sh}$, the observed transition frequency will
be dependent on $T$ \cite{tai09}. With a control of the shift to 1\%
under typical conditions we can achieve $\left|\Delta/\Omega_0
\right| < 0.01$ to 0.1.

Under the condition $|\Delta/\Omega_0| \ll 1$ we use the parabolic
approximation in Eq.~(\ref{eq:n_e_d}) (on $|T\delta|\ll 1$) to
find the leading approximation of $\overline{\delta\omega}_0$:
\begin{equation}
\overline{\delta\omega}_0\approx-\frac{1}{T}\frac{a^{(1)}}{2a^{(2)}}  \, .
\label{eq:d0}
\end{equation}
With Eq.~(\ref{eq:Aj}), the dominant dependence of
$\overline{\delta\omega}_0$ on $\Delta / \Omega_0$ can be
identified. For the usual Ramsey scheme with $\tau_1
\Omega_0$$=$$\tau_2 \Omega_0$$=$$\pi/2$ we find the expected linear
dependence:
\begin{eqnarray}
\overline{\delta\omega}_0 \approx
-\frac{1}{T}\frac{{\cal A}_{1}^{(1)}}{2{\cal A}_{0}^{(2)}}\left(\frac{\Delta}{\Omega_0}\right)
= \frac{1}{T} \frac{2 \Omega_0 T}{2 + \Omega_0 T} \frac{\Delta}{\Omega_0}.
\label{eq:d0_1}
\end{eqnarray}
Explicit analytical calculations for the first term ${\cal
A}_1^{(1)}$ in the expansion of $a^{(1)}$ of Eq.~(\ref{eq:Aj}) show
that
\begin{equation}
{\cal A}^{(1)}_1 \propto \sin [\Omega_0(\tau_1+\tau_2)/2]\,.
\label{eq:cA1}
\end{equation}
Thus, we find that ${\cal A}^{(1)}_1 = 0$ for
\begin{eqnarray}
\Omega_0(\tau_1+\tau_2) = 2 \pi n \qquad (n=1,2,3,...) \, .
\label{eq:S12}
\end{eqnarray}
From Eqs.~(\ref{eq:Aj}) and (\ref{eq:d0}), we see that for ${\cal
A}^{(1)}_1 = 0$ and ${\cal A}_{0}^{(2)}\neq 0$, the dominating
dependence of $\overline{\delta\omega}_0$ on $|\Delta / \Omega_0|\ll
1$ is now cubic:
\begin{eqnarray}
\overline{\delta\omega}_0 \approx -\frac{1}{T}\frac{{\cal A}_{3}^{(1)}}{2{\cal A}_{0}^{(2)}}\left(\frac{\Delta}{\Omega_0}\right)^3 .
\label{eq:d0_3}
\end{eqnarray}
Equation (\ref{eq:d0_3}) has two important consequences.  First,
under the condition $|\Delta/\Omega_0| \ll 1$, the resulting shift
of the central hyper-Ramsey fringe is much smaller than for the
usual Ramsey scheme (Eq.~(\ref{eq:d0_1})). Second, it has a
higher-order dependence, which is the reason for referring to this
method as the ``hyper-Ramsey" method.

\begin{figure}[t]
\centerline{\scalebox{0.7}{\includegraphics{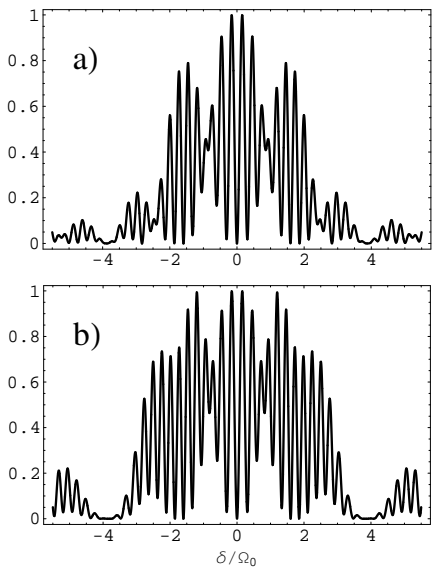}}}\caption{Hyper-Ramsey fringes $n_\mr{e}$($\delta$) under the conditions $\Omega_0\tau_1=\pi /2$, $\tau_2/\tau_1=3$, $\Omega_0T=20$, $\Delta=0$ according to a) Eq.~(\ref{eq:n_e}) and b) Eq.~(\ref{eq:hr2}).} \label{HR12}
\end{figure}

\begin{figure}[t]
\footnotesize
\includegraphics[width=8cm]{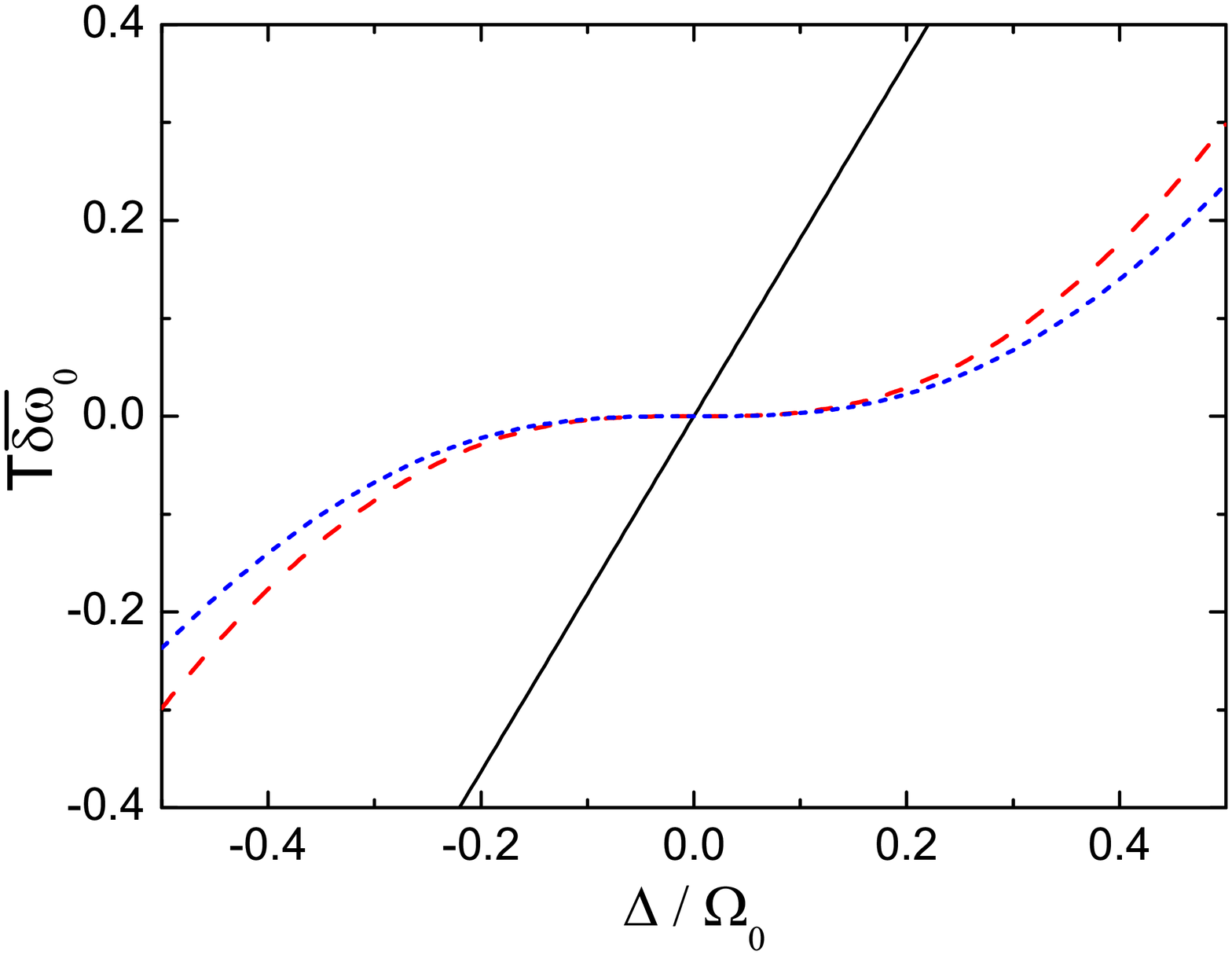}
\caption{(color) Numerically calculated shift of the central resonance $T\overline{\delta\omega}_0$ versus $\Delta / \Omega_0$ for standard Ramsey spectroscopy ($\Omega_0 \tau_1$$ = $$\Omega_0 \tau_2$$ =$$ \pi/2$; { $\Omega_0 T $$=$20}; full line) and the hyper-Ramsey method ($\Omega_0 (\tau_1 + \tau_2)$$ = $$2\pi$; $\tau_2 / \tau_1$$=$3; {$\Omega_0 T$$ =$20}). Dashed line: position of maximum; dotted line: estimate of center from signal comparison with $\pm \pi/2$ phase steps \cite{mor89}.}
\label{fig:hyper1}
\end{figure}

Apart from the condition Eq.~(\ref{eq:S12}), which essentially
minimizes the shift $\overline{\delta\omega}_0$, for applications in
spectroscopy and optical clocks it is also desirable to maximize the
amplitude of the central resonance. This is accomplished by the
maximization of the coefficient ${\cal A}^{(2)}_0$ in
Eq.~(\ref{eq:Aj}), which defines the curvature of the central
resonance top if $|\Delta / \Omega_0| \ll 1$. Under the condition of
Eq.~(\ref{eq:S12}) we find
%\begin{eqnarray}
    ${\cal A}^{(2)}_0=0.25\sin^2(\Omega_0\tau_1)$\,.
%\label{eq:cA2}
%\end{eqnarray}
Then, the coefficient ${\cal A}^{(2)}_0$ reaches its maximum at
\begin{eqnarray}
\Omega_0\tau_1=\pi (2m+1)/2 \qquad (m=0,1,2,...)\,.
\label{eq:maxA2}
\end{eqnarray}
Equations~(\ref{eq:S12}) and (\ref{eq:maxA2}) lead to the following
relationship for the values $\tau_1$ and $\tau_2$:
\begin{eqnarray}
\tau_2/\tau_1=(4n-2m-1)/(2m+1)\,.
\label{eq:t12}
\end{eqnarray}
In the simplest case $n = 1$ (i.e., when $\Omega_0(\tau_1+\tau_2) = 2\pi$) we find that either $\tau_2 / \tau_1 = 3$ or $\tau_2 / \tau_1 = 1/3$. The signal contrast is close to the maximum value of 1 (see Fig.~\ref{HR12}a). The coefficient of the cubic term of Eq.~(\ref{eq:d0_3}) in the case of $T$$\gg$$ (\tau_1+\tau_2)$ then amounts to
${\cal A}_{3}^{(1)}/2{\cal A}_{0}^{(2)}$$ \approx$$-\pi$.

The elimination of the shift $\Delta$ from the observed resonance
shift $\overline{\delta\omega}_0$ is illustrated in
Fig.~\ref{fig:hyper1} through comparison with the regular Ramsey
method using Eq.~(\ref{eq:n_e}). For the hyper-Ramsey case, we find
the residual frequency shift in two ways: by locating the maximum of
the fringe and by generating a discriminator slope at zero detuning
$\delta$ by stepping the phase of one of the pulses  by $\pm \pi/2$
in an alternating way \cite{mor89} and equalizing these signals. The
latter approach is of greater relevance for clocks, because it
directly generates an error signal with high sensitivity.  However, probing the interference at the half width points
introduces a weak linear (on $\Delta/\Omega_0$) contribution for
$\overline{\delta\omega}_0$ in the hyper-Ramsey signal.
\begin{figure}[t]
\footnotesize
\includegraphics[width=8cm]{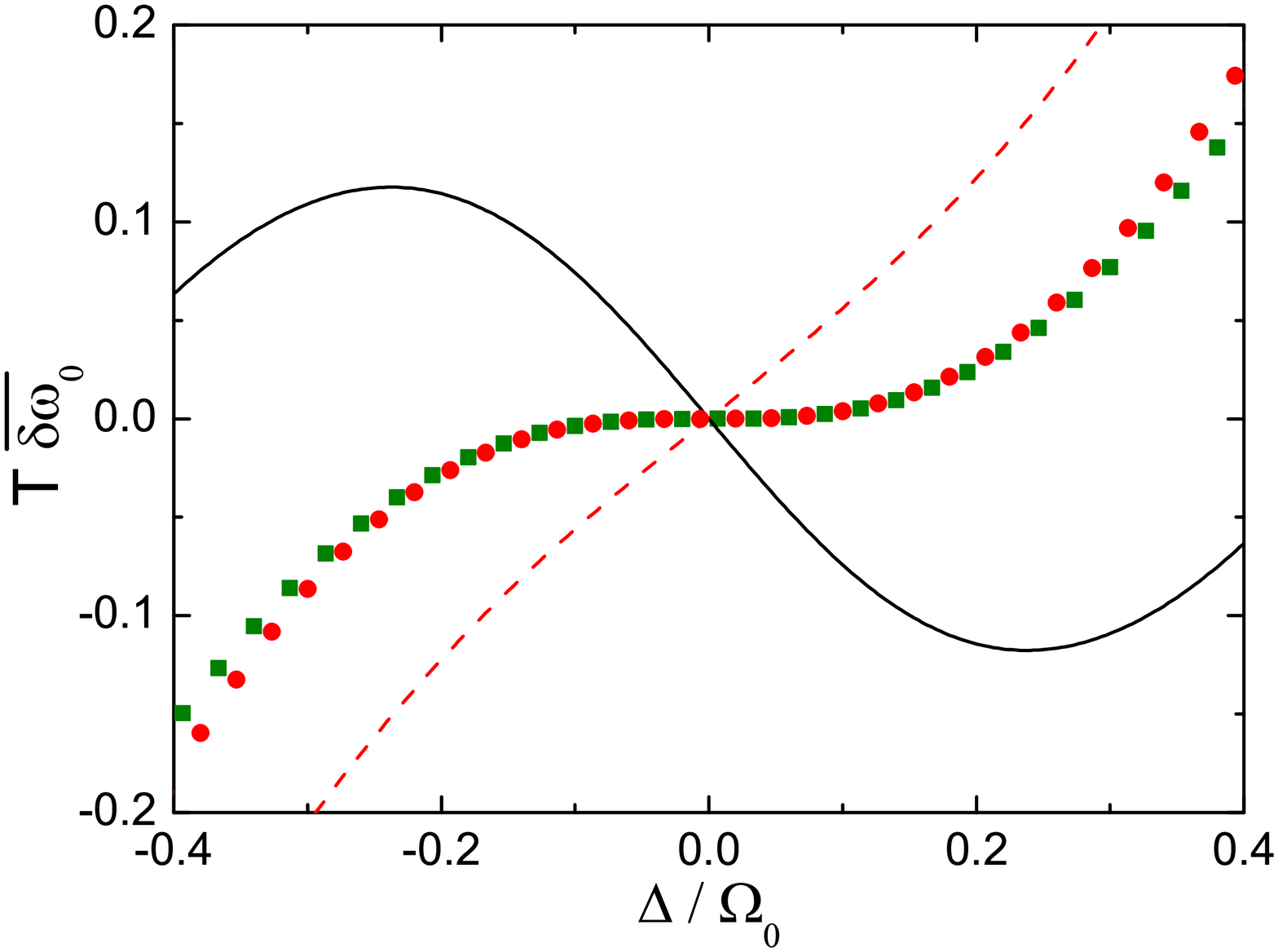}
\caption{(color) Influence of excitation pulse area on
shift suppression (in calculations we used the approach \cite{mor89} to find the center from signal comparison with $\pm \pi/2$ phase steps for the second Ramsey pulse).
Lines (solid and dotted) show numerically calculated maximum and
minimum shifts $T\overline{\delta\omega}_0$ of the central
fringe versus $\Delta /
\Omega_0$ (the normalized effective frequency shift) for the
hyper-Ramsey excitation in Fig.~\ref{fig:scheme}a for a range of the
pulse area parameter, $\Omega_0\tau_1=q\pi/2$ ($\tau_2 /
\tau_1 = 3$; $\Omega_0 T = 20$) of $0.9 \leq q \leq 1.1$.  These
lines show how the hyper-Ramsey suppression effect is compromised
by non-optimized pulse areas. The symbols (squares and circles)
show results for the hyper-Ramsey scheme in Fig.~\ref{fig:scheme}b (i.e. with additional $\pi$ phase
jumps) also for $\Omega_0\tau =q\pi/2$
($\Omega_0 T = 20$) and $0.9 \leq q \leq 1.1$. In this case the
suppression is largely insensitive to total pulse area, making
this technique more feasible experimentally.} \label{fig:hyper2}
\end{figure}

In some cases such as that for lattice-based and single
ion clocks, we cannot perfectly fulfill condition Eq.~(\ref{eq:S12}) even with good intensity control, e.g., due to
vibrational state-dependent Rabi frequencies. In
Fig.~\ref{fig:hyper2} we plot the resultant
$\overline{\delta\omega}_0$ versus ($\Delta/\Omega_0$) dependence
for non-optimal values of the sum $\tau_1 + \tau_2$.  Here
we see how a deviation from $\Omega_0( \tau_1 + \tau_2) = 2\pi$
re-introduces the linear dependence on $\Delta / \Omega_0$, though
still with a strongly reduced amplitude compared to that for usual
Ramsey spectroscopy.% (Fig.~\ref{fig:hyper1}).

However, this inherent problem can be easily overcome by introducing
a phase jump of $\pi$ at the beginning of the second pulse of the
hyper-Ramsey sequence (Fig.~\ref{fig:scheme}b). The phase is stepped
back after $2/3$ of the second pulse, such that the second Ramsey
pulse can be seen as being composed of a pulse with $-\Omega_0$
(echo pulse) directly followed by one with $\Omega_0$. With modern
electronic oscillators, namely direct digital synthesizers, it is
easily possible to maintain phase coherence over both frequency
steps and phase jumps. Note, the second pulse for the excitation in
Fig.~\ref{fig:scheme}b has a technical similarity to the composite
pulses used in NMR spectroscopy \cite{lev96a}.

The expression for the population $n_\mr{e}$ can be extended to
include the phase step in Fig.~\ref{fig:scheme}b and we then find:
\begin{eqnarray}
\label{eq:hr2}
&& n_\mr{e}=\\
&&\left|\langle e|\widehat{W}(\tau,\Omega_0,\delta_\mr{p})\widehat{W}(2\tau,-\Omega_0,\delta_\mr{p})\widehat{V}(T\delta) \widehat{W}(\tau,\Omega_0,\delta_\mr{p})|g\rangle \right|^2,\nonumber
\end{eqnarray}
where $\tau$ is the duration of the short pulse. In this case ${\cal
A}^{(1)}_{1}=0$, i.e., the cubic dependence of Eq.~(\ref{eq:d0_3})
applies for arbitrary values $\Omega_0$ and $\tau$. Under conditions
$\Omega_0\tau =\pi / 2$ and $T \gg 4\tau$ we then find that ${\cal
A}_{3}^{(1)}/2{\cal A}_{0}^{(2)}$$\approx$$-4$.  The curves in
Fig.~\ref{fig:hyper2} were calculated from Eq.~(\ref{eq:hr2}) using
$\Omega_0\tau \approx \pi / 2$ to maximize ($\sim$1) the signal contrast (see
Fig.~\ref{HR12}b). They show the large advantage of adding the phase
step to the hyper-Ramsey sequence in terms of suppressing the
dependence on pulse area.

Let us consider a numerical example using magnetically induced
spectroscopy of $^{174}$Yb \cite{bar06,tai06}. Since the advantage
of the hyper-Ramsey method over usual Ramsey spectroscopy is
obvious (by 2-4 orders for $\left|\Delta/\Omega_0
\right|$$<$$0.1$-0.01), here we compare hyper-Ramsey to Rabi
spectroscopy. If we assume typical experimental conditions of $T
= 40$~ms, $\tau =10$~ms ($\Omega_0 / 2 \pi =25$~Hz) at a magnetic
field of 2~mT, an ac-Stark shift of 70~Hz (during pulses) will
result. If we further assume that we can control the intensity to
1\%, we should be able to zero the effective detuning with an
uncertainty of $\Delta/2\pi$$\approx$~$0.7$~Hz. The resulting
suppression inherent in the technique then leaves a shift (and
resultant uncertainty) of 0.35~mHz (the fractional level is below of
10$^{-18}$ for 518 THz clock transition) for the central fringe,
which is a factor of 120 less than the uncertainty for the
corresponding Rabi case (with 1\% control on a 4.2~Hz shift for a
80~ms $\pi$-pulse).

Thus, the hyper-Ramsey method is a novel technique that offers a spectroscopic signal that is virtually free from excitation-induced frequency shifts and their fluctuations. It is robust against perturbations and noise, because the influence of the shifts on the signal is essentially eliminated. This method will now allow a variety of systems to enjoy other advantages of Ramsey spectroscopy: increased resolution for a given interrogation time, and improved stability \cite{dic87,que03}. The hyper-Ramsey method has broad applications for optical clocks, especially those based on ultra-narrow transitions, two-photon transitions, or lattice clocks based on bosonic isotopes with controlled collision shifts \cite{aka08,lis09}. Moreover, our approach opens a prospect for the high-precision optical clocks based on direct frequency comb spectroscopy. High resolution matter-wave sensors \cite{yve03} are also expected to benefit from the suppression of phase shifts in the interference patterns due to the excitation pulses.

We thank Ch.~Tamm, E.~Peik, T.~Mehlst\"{a}ubler and T.~Heavner for useful discussions. A.V.T. and V.I.Yu. were supported by RFBR (08-02-01108, 10-02-00591, 10-08-00844) and programs of RAS.
V.I.Yu., U.S., Ch.L. and F.R. acknowledge gratefully support by
the Center for Quantum Engineering and Space-Time Research
(QUEST), the European Community's ERA-NET-Plus Programme under
Grant Agreement No.~217257, and by the ESA and DLR in the project
Space Optical Clocks.

V.I.Yudin e-mail address: viyudin@mail.ru

\end{document}